\documentclass[%
 reprint,
 amsmath,amssymb,
 aps,
]{revtex4-2}

\usepackage{graphicx}
\usepackage[colorlinks=true, allcolors=blue]{hyperref}
\usepackage{dcolumn}
\usepackage{bm}
\usepackage{float}
\usepackage[caption=false]{subfig}
\usepackage{flushend}
\usepackage{physics,amsthm}

\usepackage{placeins}

\begin{document}

\title{Accurate Learning of Equivariant Quantum Systems from a Single Ground State}

\author{Štěpán Šmíd}
\affiliation{%
Department of Computing, Imperial College London, London SW7 2AZ, United Kingdom
}%
\author{Roberto Bondesan}%
\affiliation{%
Department of Computing, Imperial College London, London SW7 2AZ, United Kingdom
}%

\date{\today}

\begin{abstract} 
Predicting properties across system parameters is an important task in quantum physics, with applications ranging from molecular dynamics to variational quantum algorithms. Recently, provably efficient algorithms to solve this task for ground states within a gapped phase were developed. Here we dramatically improve the efficiency of these algorithms by showing how to learn properties of all ground states for systems with periodic boundary conditions from a single ground state sample. We prove that the prediction error tends to zero in the thermodynamic limit and numerically verify the results.
\end{abstract}

\maketitle



\section{Introduction}

The simulation of quantum many-body systems is a fundamental task in quantum physics and underlies our understanding of nature. 
Finding ground states and learning their properties is essential for developments in quantum chemistry and the discovery of technologies used in our everyday lives, such as renewable energy storage \cite{zitnick2020introduction}. 
Computing ground states of classical Hamiltonians also allows us to solve classical optimization problems \cite{Optimisation_as_Ising}.
The time required for exact diagonalization however grows exponentially with the system size, which makes it practically infeasible for anything but the smallest examples. This indicates the necessity for invention of advanced methods simplifying this task while maintaining desirable accuracy. 
While many clever classical algorithms have been invented, like the quantum Monte Carlo \cite{RevModPhys.73.33} methods or the Density Matrix Renormalization Group (DMRG) \cite{Schollw_ck_2011}, we are still limited by our computing power.
Further, despite the existence of quantum algorithms with provable advantage for quantum simulations, the state-of-the-art quantum computers are still too noisy and imperfect to implement these.

Recently, building on the success of machine learning (ML) for image and text processing, 
researchers have used classical ML techniques to predict properties of parameterized Hamiltonians $H(x)$, where $x \in [-1,1]^m$ are parameters, by training on data for distinct choices of $x$. This is the setting for example when predicting molecular properties in a new atomic configuration  based on knowledge of previous configurations \cite{Batzner_2022,satorras2022en}. 
These techniques can also be applied in the context of variational quantum algorithms to predict observables at new variational parameters, allowing us to reduce the cost of running quantum computations \cite{sung2020using,nicoli2023physicsinformed}. 
It was previously shown that for a system with short-range interactions the ground state expectation values of 
observables which are sum of geometrically local terms can be 
accurately approximated within a gapped phase using 
a number of samples scaling only logarithmically with the system size \cite{lewis2023improved,onorati2023efficient}.

In \cite{smid2023efficient}, similar guarantees and efficient scaling with the number of qubits were obtained for systems with long-range interactions -- specifically decaying either exponentially or as power law with decay exponent greater than twice the dimension of the system. In \cite{smid2023efficient}, a further complexity reduction was shown using the tools of equivariant ML. Assuming the ML model used to be equivariant under the automorphism group $G$ of the interaction hypergraph (i.e.~the hypergraph with the same vertex set as the lattice, and with one hyperedge per interaction), we can use the same model for predicting the few-body constituent terms $O_I$ of an observable $\sum_I O_I$ across the whole system. This reduces the complexity by a factor of $|G|$, which for periodic systems where $|G|$ grows linearly with system size means that a constant number of training samples is sufficient for a constant prediction error. Note that this applies beyond translation invariant systems.

In this Letter, we shall present a surprising application of these latest developments. We will prove that for a large class of equivariant systems knowledge of the ground state for a single choice of the parameters $x$ is enough to 
learn the whole topological phase and predict ground state properties for any other parameter choice within it.
Since obtaining the data needed for training the models is costly and preparing ground states for many different parameter choices can be experimentally challenging, this conceptually novel approach presents clear practical benefits. Following the idea that in an equivariant quantum system we may use the same ML model trained on predicting a $p$-body observable to predict all such observables across the system using its symmetries, here we investigate the idea of also using different parts of the same system as distinct training samples.

Given that the cardinality of the automorphism group $G$ of the systems considered 
grows with their size, we can use the theoretical guarantees of \cite{smid2023efficient,lewis2023improved} to show that the prediction error of this method tends to $0$ as the system size increases to infinity. \textit{This tells us that a single ground state of such equivariant systems carries not only enough information to predict properties across the whole gapped phase, but also does so in a precise manner in the thermodynamic limit}. We  further specify an asymptotic upper bound for the average error of this method for systems with periodic boundary conditions in any number of dimensions. We verify these predictions with DMRG simulations of disordered Heisenberg models and long-range Ising models in $1$D, which can be realized experimentally in trapped ions \cite{hauke2010complete} and polar molecules \cite{yan2013observation}.
\FloatBarrier
\pagebreak

\onecolumngrid

\begin{figure}[H]
    \centering
    \includegraphics[width=\textwidth]{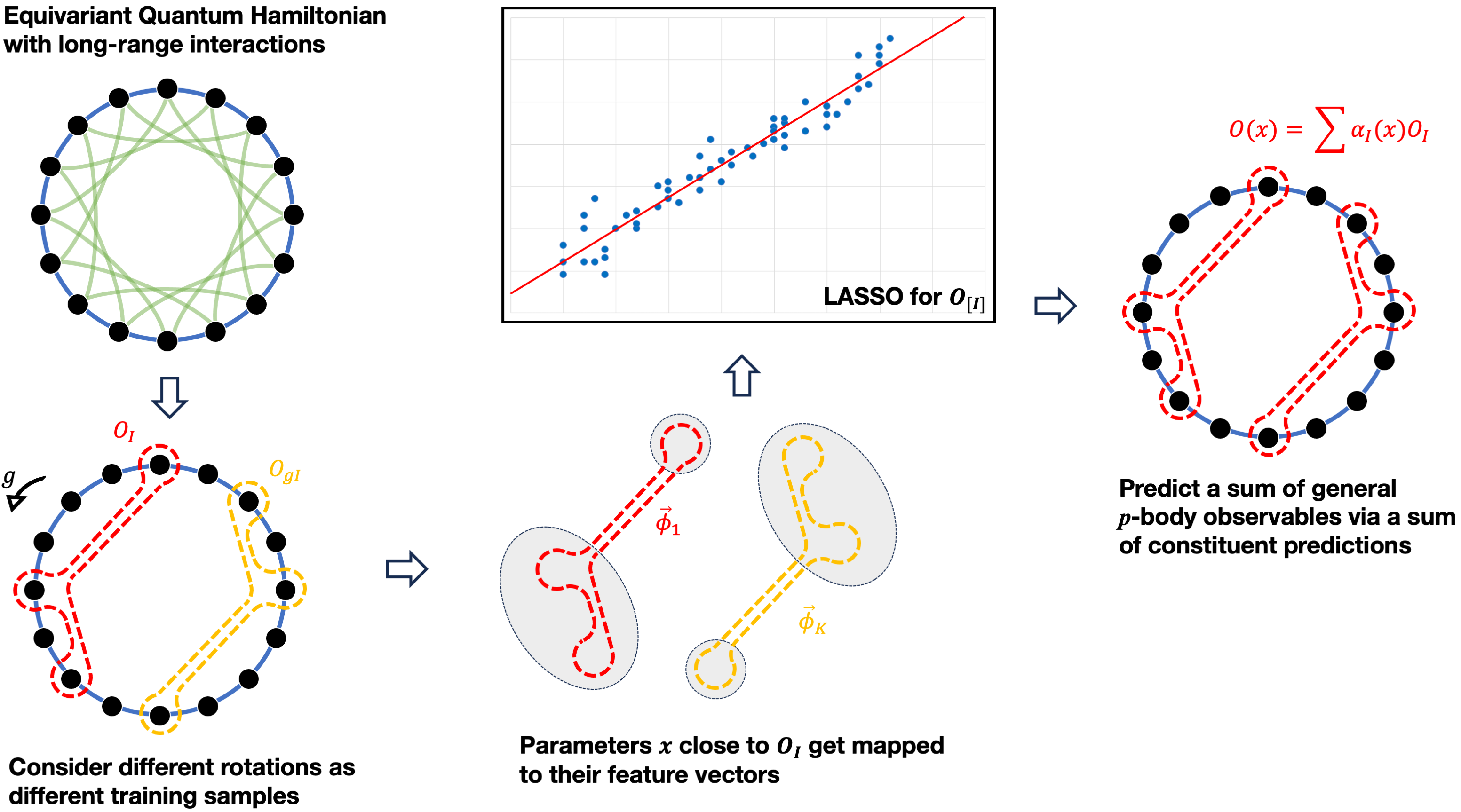}
    \caption{\label{fig:summary}\textbf{Summary of the algorithm.} Given a representation of a single ground state of an equivariant Hamiltonian with long-range interactions, we would use its symmetries to think of different parts of the ground state as distinct training samples for a machine learning model trained for predicting a $p$-body observable. This way we can train a machine learning model for each distinct kind of a constituent term in a sum of $p$-body observables from a single ground state sample, and hence predict the sum itself for any other ground state within the topological phase.}
\end{figure}
\twocolumngrid

\section{The algorithm}

We look at the problem of learning the ground state expectation values of observables across Hamiltonian parameters for a system of qubits on a lattice with vertex set $V$.
The important observation in this Letter is that the classical data sets required for training the machine learning models may be obtained from many different measurements of a single ground state sample by exploiting the symmetries of the system. 

We consider parameters $x\in [-1,1]^m$ and 
Hamiltonians $H(x)=\sum_{I\in \mathcal{E}}h_I(x_I)$, 
so that $x=(x_I)_{I\in {\cal E}}$.
Here $\mathcal{E}\subseteq \mathcal{P}_k$ with $\mathcal{P}_k$ the set of subsets of lattice vertices with at most $k$ elements, $k=\mathcal{O}(1)$. 
We consider predicting the expectation value 
in the ground state $\rho(x)$ 
of any sum of $p$-body observables \begin{equation}
\label{observable} 
O(x) = \sum_{I\in \mathcal{S}} \alpha_I(x) O_I,
\end{equation} 
where $\alpha_I(x)$ are coefficients and $\mathcal{S}\subseteq \mathcal{P}_p$ with $p=\mathcal{O}(1)$. 
We also denote 
$f(O,x) = \Tr(O \rho(x))$.

The observable $O(x)$ is a priori assumed to have a bounded $\ell_1$ norm, $\| O \|_1 = \mathcal{O}(1)$, but later we shall use a super-normalized version to prevent the expectation values from getting polynomially small (see \cite{smid2023efficient} for details). 

We define the interaction hypergraph as the triple $(V,\mathcal{E},h)$, where $h$ is a function that assigns to each $I\in \mathcal{E}$ the interaction term $h_I(x_I)$.
We denote by $G$ the automorphism group of the interaction hypergraph.
This is the set of permutations of vertices that permute the interactions terms; specifically, $g\in G$ if
\begin{align}
    U_g h_I(x_I) U_g^{-1}
    =
    h_{gI}(x_I)\,,
\end{align}
where $U_g$ is the unitary representation of a permutation $g$ on the Hilbert space of the quantum theory and $gI$ is the permuted set of vertices.
This implies that the Hamiltonian and the ground state are $G$-equivariant: 
\begin{align}
U_g \rho(x) U_g^{-1}=\rho(g^{-1}\cdot x)    
\end{align}
where
$g\cdot x = (x_{gI})_{I\in {\cal E}}$.
For example, for a system with periodic boundary conditions and pairwise interactions
$h_{ij}(x_{ij})=x_{ij}f(d(i,j)) H_{ij}$ with $d(i,j)$ the Euclidean distance and $f$ an arbitrary function, $G$ is the Euclidean group of symmetries of the lattice under which $d(i,j)$ is invariant, and 
$|G| = \Theta(n)$ \footnote{We recall that $f(n) = \Theta(g(n))$ if there are $c_1,c_2>0$ and $n_0$
such that $c_1 |g(n)| \le |f(n)| \le c_2 |g(n)|$ for all $n \ge n_0$}.
This setting includes nearest neighbor and long-range interactions.


We now consider $\mathcal{S}/G$, the set of equivalence classes $[I]$, where $[I]$ is the orbit of $I$ under $G$.
We then produce a dataset 
\begin{equation} 
S_{[I]} = \left\{g \cdot x_0, f(O_{g I},x_0)\right\}_{g \in G},
\end{equation}
for each $[I]\in \mathcal{S}/G$
from the single ground state $\rho(x_0)$.
Using the equivariance of expectation values 
\begin{align}
    \label{eq:equiv_f}
    f(O_{gI},x)
    =
    \Tr(O_IU_g^{-1}
    \rho(x)U_g)
    =
    f(O_{I},g\cdot x)\,,
\end{align}
$S_{[I]}$ corresponds to the expectation values of $O_I$ at $|G|$ distinct points in parameter space.
$S_{[I]}$ is a set of i.i.d.~samples if $\{g\cdot x_0\}_{g\in G}$ are i.i.d.
This happens if the distribution from which $x_0$ is sampled is $G$-invariant, and is such that the $\{g\cdot x_0\}_{g\in G}$ are uncorrelated. 
In the experimental section we will consider a uniform distribution over $[-1,1]^m$
which satisfies these conditions.
We now use $S_{[I]}$ to train a ML model 
$\hat{f}(O_I,x)$ to predict $f(O_I,x)$ for all $x$.
Note that by making $\hat{f}$ equivariant as in \eqref{eq:equiv_f} we only need $|\mathcal{S}/G|$ different ML models to predict $O(x)$ across all $x$.
Figure \ref{fig:summary} summarizes the approach and uses the ML algorithm presented in \cite{lewis2023improved,smid2023efficient}.
This algorithm is suitable for short-range correlated phases, where we map the parameters $x$ in the neighborhood of $O_I$ to their random Fourier feature vectors $\vec{\phi}$, which are then used to train LASSO, an $\ell_1$-regularized linear regression.

The classical data sets $S_{[I]}$ needed for  training may be obtained from some classical representation of the ground state $\rho(x_0)$, like a matrix product state or a classical shadow \cite{Huang_2020}. They may be also obtained from a single copy of the quantum state using quantum ground state restoration \cite{QuantumGroundStateRestoration}, although in practice, preparing the same ground state repeatedly and measuring different observables is currently much more feasible.

The method presented is most suitable for
predicting observables in systems with an extensive number of parameters such as disordered systems. It is not applicable for example in the case where we would have a global parameter which should differ across different samples, as there is clearly no way of learning its dependencies from a single sample.

This method may also be  used for predicting the classical shadows of ground states $\rho(x)$, or in fact any other quantum tomography technique, instead of just a single observable $O(x)$.
In particular, we can $\epsilon$-approximate all reduced density matrices over $\mathcal{O}(1)$ qubits with $T=\mathcal{O}(\log(n))$ randomized single qubit measurements of $\rho(x_0)$ via its classical shadow 
$\sigma_T(s;x_0)$ \cite{Huang_2022}, where $s=(s^1,\dots,s^T)$ are the measurement outcomes.
The classical dataset $S'_{[I]}
=\{ g\cdot x_0, \sigma_T(g\cdot s;x_0)\}_{g\in G}$ can then be used to predict all reduced density matrices over $\mathcal{O}(1)$ qubits across the parameter space.
These 
can be further used to predict any (nonlinear) analytic functions of the reduced density matrices, such as entanglement entropies, quantum fidelities, or correlation functions.

Further, we could theoretically use techniques for Hamiltonian reconstruction developed in \cite{Qi_2019}, which would allow us to use the knowledge of the ground state in order to express its parent $k$-local Hamiltonian in a given basis. This method would enable us to determine the dependence of the Hamiltonian on its parameters from a single ground state sample. 
But we found this technique to be very sensitive to errors in the values of correlations in the ground state, even so that the simulations of larger systems themselves do not always lead to a faithful Hamiltonian reconstruction, and hence is not practically usable for reconstruction from predictions.

\section{Error bounds and complexity}

In this section we discuss the performance of the algorithm above by deriving rigorous error bounds and runtime complexity.
We focus on  gapped Hamiltonians with short-range correlations for which rigorous sample complexity bounds exist \cite{lewis2023improved,onorati2023efficient,smid2023efficient}.
We compute the error of our procedure to predict $\Tr(O_I \rho(x))$ with
$O_I$ supported on
$\mathcal{O}(1)$ qubits.
The number of samples $N$  to  predict this observable across a gapped phase with generalization error $\epsilon$ is 
independent of the number of qubits, $N=f(\epsilon)$.
Our algorithm has $N=|G|$ and achieves error \footnote{More meticulously, we have an upper bound $N = \mathcal{O}(f(\epsilon))$ on the number of samples needed to achieve a given error, and hence by providing asymptotically more samples we can guarantee obtaining such an error.}
\begin{align}
    \epsilon = f^{-1}(|G|)\,.
\end{align}
Now we use the explicit expressions for the sample complexity of \cite{smid2023efficient}:
\begin{align}
    \label{eq:f_eps}
    f(\epsilon)
    =
    \begin{cases}
    2^{\operatorname{polylog}(1/\epsilon)} & \textup{ exp decay/short range},\\
    2^{\mathcal{O}(\epsilon^{-\omega} \log(1/\epsilon))} & \textup{ power law, } \alpha>2D  \,.
    \end{cases}
\end{align}
Here $\alpha$ is the power-law exponent, $D$ the system's dimension, and $\omega$ is a function of $D$ and the strength of the interactions, whose specific form is given in Appendix \ref{sec:omega dependence}, and which diverges as $\alpha \to 2D$.

We see from \eqref{eq:f_eps}
that for exponentially decaying  or short range interactions $f(\epsilon) = 2^{-c \log^d(\epsilon)} \stackrel{\text{set}}{=} c'n$ for some positive $c,c'$ and integer $d$,
and  the error of our procedure scales as:
\begin{equation}\label{eqn:error bound short} \epsilon = 2^{-\mathcal{O}\left(\sqrt[d]{\log(n)}\right )}\,.
\end{equation} 
Deriving the error for power law interactions is more involved, and is derived in detail in Appendix \ref{sec:error derivation}, giving
\begin{equation}\label{eqn:error bound power law}\epsilon = \mathcal{O}\left(\sqrt[\omega]{\frac{\log(\omega\log(n))}{\omega \log(n)}}\right)\,.
\end{equation}

Note that in both cases $\epsilon\to 0$ as $n\to\infty$, proving that this algorithm achieves small error for large enough systems. The same error bounds also apply when predicting the reduced density matrices via classical shadows.

In order to predict an observable $O(x)$ as specified previously, we would need $|\mathcal{S}/G|$ different ML models, each one with a constant number of features, and so the total training time would be $\mathcal{O}(|\mathcal{S}/G| \cdot N)$, while the prediction time would be $\mathcal{O}(|\mathcal{S}|)$, making the overall run time of this algorithm $\mathcal{O}(|\mathcal{S}/G| \cdot N + |\mathcal{S}|) = \mathcal{O}(|\mathcal{S}|)$, as $N \leq |G|$. This in particular means that taking less than $|G|$ training samples does not decrease the time complexity, as the prediction time dominates, and so there is no penalty when focusing on maximal accuracy.

\section{Simulations}

The data for large systems with up to $128$ sites on a periodic chain were obtained using the density matrix renormalization group method 
\cite{Schollw_ck_2011}. The details of the numerical implementations are specified in \cite{smid2023efficient} and the corresponding code is also available at \cite{Smid_Efficient_Learning_of_2024}.

The Hamiltonian for the Heisenberg model considered is given by \begin{equation}\label{eqn:Heisenberg}
H = \sum\limits_{i=0}^{n-1} J_{i,i+1} (X_i X_{i+1} + Y_i Y_{i+1} + Z_i Z_{i+1})\, ,   
\end{equation}
where $i+1$ is understood modulo $n$. Here we sample the $J_{i,i+1}$'s uniformly at random from $[0,2]$.

\begin{figure}[H]
    \centering
   \subfloat[GS energy in Heisenberg model \eqref{eqn:Heisenberg}]{\label{fig:Plot-Heisenberg}\includegraphics[width = 0.33\textwidth]{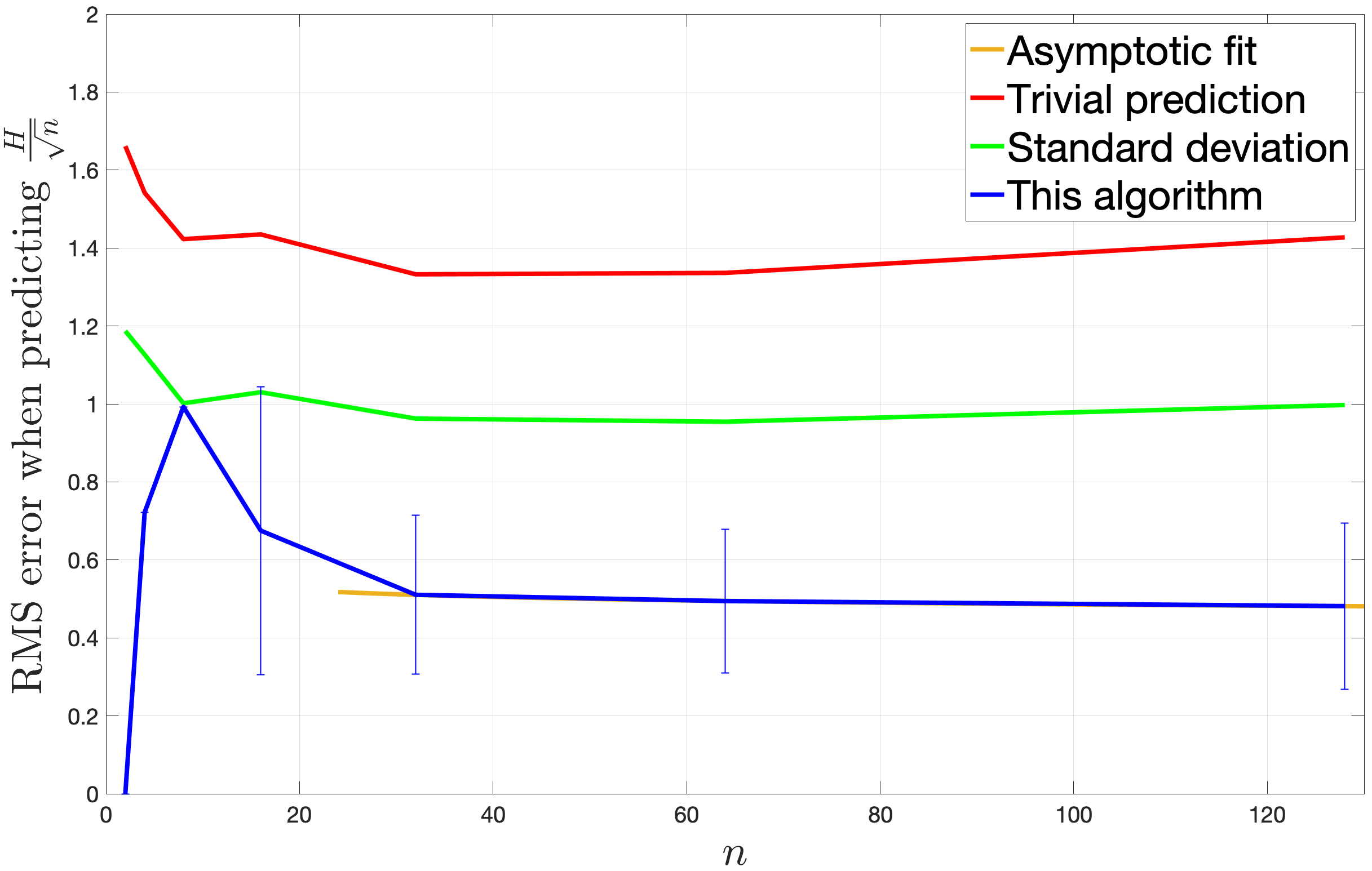}}
    \subfloat[All $C_{ij}$ in Heisenberg model \eqref{eqn:Heisenberg}]{\label{fig:Plot-Heisenberg-AllCorrsDecay}\includegraphics[width = 0.33\textwidth]{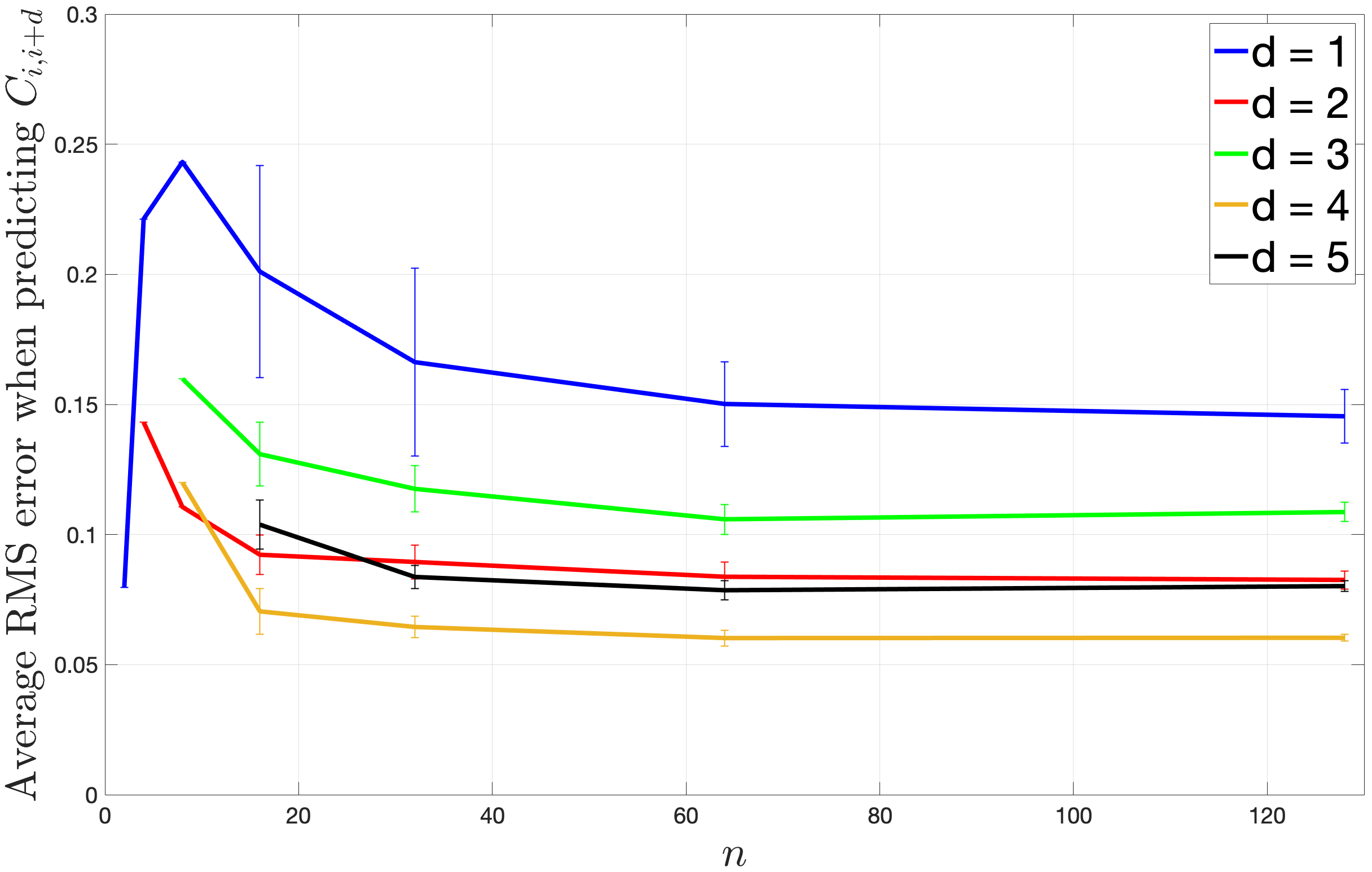}}
    \subfloat[GS energy in Ising model \eqref{eqn:Ising}]{\label{fig:Plot-Ising}\includegraphics[width = 0.33\textwidth]{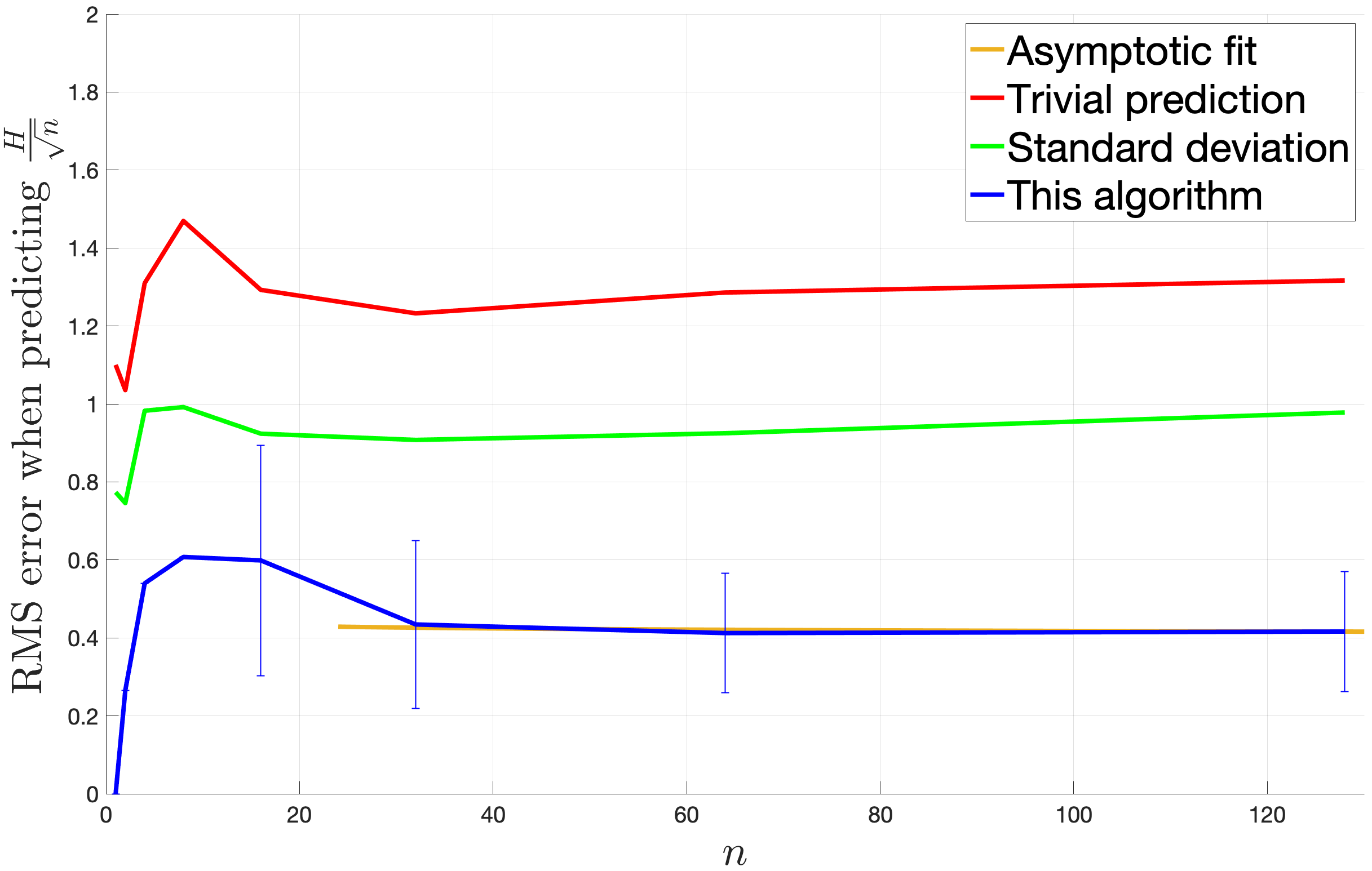}}
    \caption{\textbf{Numerical results.} \textbf{(a) and (c)} Comparing the average root-mean-square error obtained with our method to the trivial predictor and the standard deviation of the testing samples, when predicting the ground state energy $O = H/\sqrt{n}$ as the system size increases. 
    \textbf{(b)} The average RMS error obtained when predicting all $C_{ij} = \frac{1}{3}(X_i X_j + Y_i Y_j + Z_i Z_j)$ in the Heisenberg model using classical shadows, showing different distances of the two qubits separately, as the system size increases.}
\end{figure}

The results of the average root-mean-square error obtained with this method when predicting the normalized ground state energy $O = \frac{H}{\sqrt{n}}$ of this Heisenberg model are shown on Figure \ref{fig:Plot-Heisenberg}. We compare these to the trivial predictor, which always outputs the energy value of the single training sample, as well as to the standard deviation of the testing samples. One can clearly see that the error does indeed decrease with system size, and at $n = 128$ sites is already well below half a standard deviation. This non-obvious normalization of the energy ensures asymptotically constant standard deviation of the samples, which is in order to counteract the effects stemming from the central limit theorem applied to the random variables $\Tr(O_I\rho(x))$. 
The reasons behind this and the applicability of these ML techniques to such observables are discussed in detail in \cite{smid2023efficient}.

\begin{figure}[H]
    \centering
    \bigskip
    \includegraphics[width=0.42\textwidth]{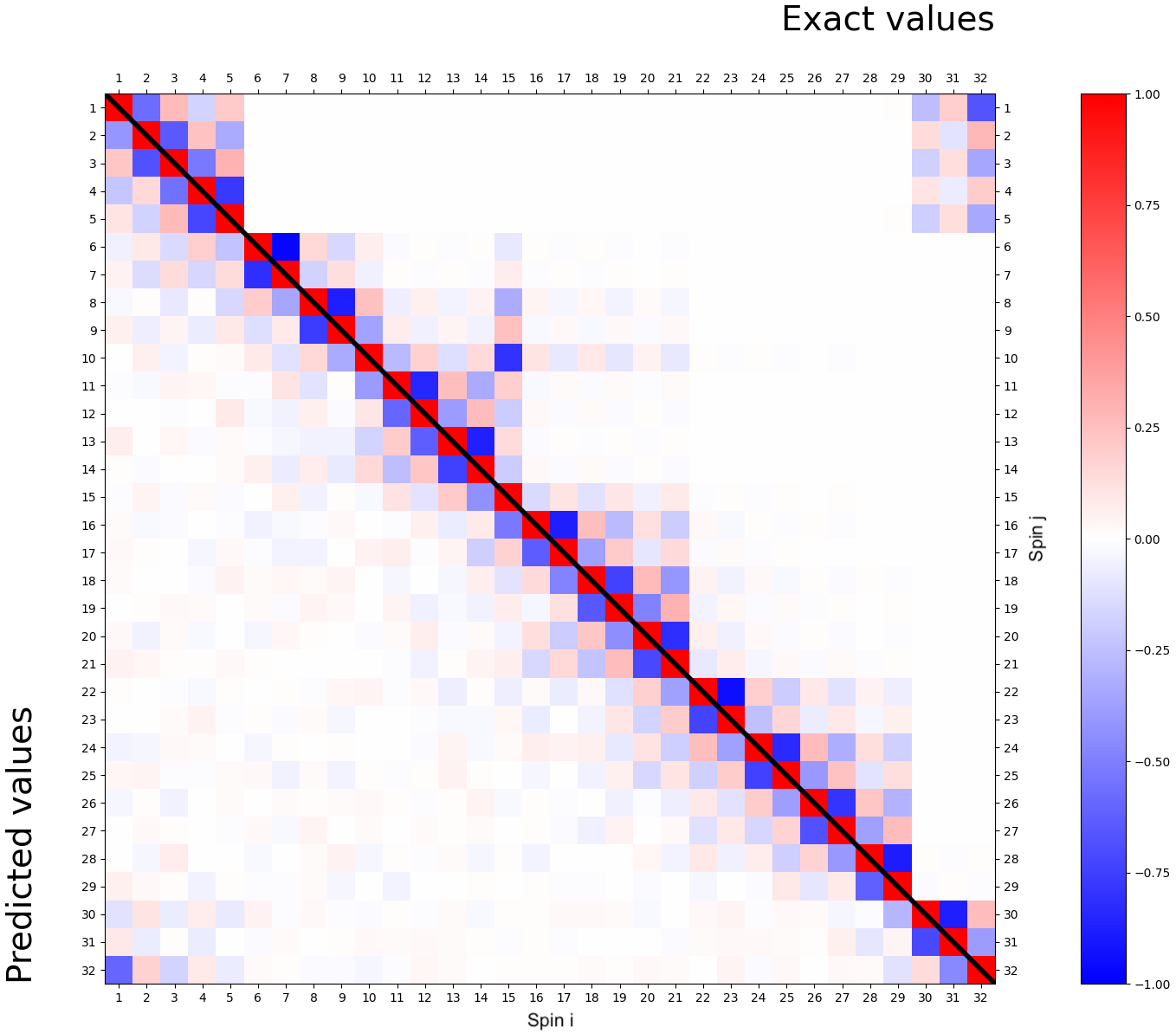}
    \caption{\textbf{Classical shadows.} Plotting an instance of all correlations $C_{ij}$ in a $32$ qubit Heisenberg chain, comparing the exact values above the diagonal to the predicted values below the diagonal.}
    \label{fig:Heisenberg-AllCorrs}
\end{figure}

In Figures \ref{fig:Heisenberg-AllCorrs} and \ref{fig:Plot-Heisenberg-AllCorrsDecay} we look at the classical shadows of the ground states in the Heisenberg model. 
We first do $T = \mathcal{O}(\log(n))$ complete measurements of the training ground state in random Pauli bases, which gives us the classical shadow representation. Then we use this representation to calculate all two-body reduced density matrices, on which we subsequently train the ML models. Finally, we use these models to predict the reduced density matrices at new parameter choices and calculate the values of all correlations $C_{ij} = \frac{1}{3}(X_i X_j + Y_i Y_j + Z_i Z_j)$. Figure  \ref{fig:Heisenberg-AllCorrs} shows an instance of such a prediction, comparing exact values obtained from DMRG (shown above the diagonal) to the values predicted with our algorithm (shown below the diagonal). Hence the accuracy of ML can be understood by the symmetry of this plot. Figure \ref{fig:Plot-Heisenberg-AllCorrsDecay} then shows the scaling of these predictions, plotting the error when predicting $C_{i,i+d}$ as the system size increases separately for different distances $d$.

The obtained error shown on Figures \ref{fig:Plot-Heisenberg} and \ref{fig:Plot-Ising} is also illustratively fitted with the predicted bounds \eqref{eqn:error bound short} and \eqref{eqn:error bound power law} respectively, demonstrating good agreement with the theory.


\onecolumngrid

\newpage
\twocolumngrid

Likewise, the Hamiltonian for the Ising model considered is given by \begin{equation}\label{eqn:Ising}H = \sum\limits_{i<j} \frac{1+J_i J_j}{d(i,j)^\alpha} Z_i  Z_j + \sum\limits_{i=0}^{n-1} h_i  X_i\, ,\end{equation} where $d(i,j) = \min\{|i-j|,n-|i-j|\}$ is the metric on a periodic chain. Here we sample the $J_i$'s uniformly at random from $[0,2]$, while the $h_i$'s are sampled uniformly at random from $[0,e]$, and the decay exponent is taken to be $\alpha = 3$. 
The results for this model are correspondingly presented in Figure \ref{fig:Plot-Ising}, showing a similar behavior to that of short-range interactions, but with the error decreasing even slower. 
For this Ising model, $\omega$ appearing in the error bound evaluates to $\omega = \frac{864}{83} \approx 10.41$. 

\section{Conclusion}

We presented a novel approach to 
predict ground states across a whole topological phase from the knowledge of a single ground state.
We have proven a decreasing upper bound of the average prediction error for systems with periodic boundary conditions, and presented numerical simulations of $1$D chains with either short-range or long-range interactions, demonstrating our findings in practice.
This research allows one to dramatically reduce the resources needed to characterise a topological phase. Future directions include extensions to gapless phases and Coulomb interactions.

\begin{acknowledgments}
ŠŠ acknowledges funding by a PhD scholarship from the Department of Computing at Imperial College London.
\end{acknowledgments}

\appendix

\section{$\omega$ dependence}\label{sec:omega dependence}

For a $k$-local Hamiltonian on a $D$-dimensional lattice, whose interactions decay as a power law with exponent $\alpha >2D$, we may calculate the exponent $\omega$ appearing in the error dependence of sampling complexity using the following formulas:

\begin{equation*}
    \begin{aligned}[c]
        \omega &= \frac{k D}{\nu - D}\,,\\
        \alpha' &= \alpha-2D-\epsilon\,,\\
        \epsilon &= \frac{1}{2}\cdot \frac{(\alpha-2D)^2}{(\alpha-2D)^2+\alpha-D}\,,\\
    \end{aligned}
    \quad
    \begin{aligned}[c]
        \nu &= \alpha' (\beta(1-\eta)-\eta)\,,\\
        \beta &= \frac{\alpha-D}{\alpha-2D}-\frac{\epsilon}{2}\,,\\
        \eta &= \frac{1}{2} \cdot\frac{\alpha-2D}{2\alpha - 2D - 1}\,.
    \end{aligned}
    \medskip
\end{equation*}

\medskip
\section{Error bound derivation}\label{sec:error derivation}

Starting from the sampling complexity $$N = 2^{\mathcal{O}(\epsilon^{-\omega}\log(1/\epsilon))}\,,$$ we have that there exist $c$ and $\epsilon_0$ such that $$N \leq e^{c\epsilon^{-\omega}\log(1/\epsilon)} \triangleq f(\epsilon)$$ whenever $0<\epsilon < \epsilon_0$. Setting $f(\epsilon) = n$, we get that \begin{align*}-\omega\log(\epsilon) e^{-\omega\log(\epsilon)} = \frac{\omega}{c}\log(n)\,.\end{align*}

To solve this, we will use the principal branch of the Lambert $W$ function, denoted by $W(x)$, to get that \begin{align*} -\omega\log(\epsilon) &= W\left(\frac{\omega}{c}\log(n) \right),\\
\epsilon^{-\omega} &= e^{W\left(\frac{\omega}{c}\log(n) \right)}\,.\end{align*} Using properties of $W$, we can move it out of the exponent like \begin{align*}\epsilon^{-\omega} &= \frac{\frac{\omega}{c}\log(n) }{W\left(\frac{\omega}{c}\log(n) \right)}\,,\\
\epsilon &= \sqrt[\omega]{\frac{W\left(\frac{\omega}{c}\log(n) \right)}{\frac{\omega}{c}\log(n)} }\,.
\end{align*} Finally, we can now use that $W(x) \sim \log(x)-\log(\log(x))+o(1)$ as $x \to \infty$ to obtain $$\epsilon \sim \sqrt[\omega]{c}\sqrt[\omega]{\frac{\log\left(\omega\log(n) \right)}{\omega\log(n)} }\,.$$

Note that because of the presence of the error in the exponent of the sampling complexity, using this last form for $\epsilon$ already yields that $N \leq o(n)$, so providing $N = \Theta(n)$ is sufficient to ensure that this gives an upper bound on $\epsilon$.


\nocite{*}

\bibliography{bibliography}

\end{document}